\documentclass[aps,nofootinbib,superscriptaddress,showpacs,floatfix,preprintnumbers,twocolumn]{revtex4}
\usepackage{amssymb,amsmath,graphicx,xcolor,slashed}
\usepackage[colorlinks=true,linkcolor=blue,citecolor=blue,urlcolor=blue]{hyperref}
\usepackage{cleveref}
\usepackage{physics}
\usepackage{subfigure}
\usepackage{multirow}
\usepackage{comment}
\usepackage{lineno}

\newcommand{\LambdaQCD}{\Lambda_\text{QCD}}
\newcommand{\diff}[2]{\operatorname{d}{\hspace{-0.15em}}^{#1}{#2}\hspace{0.15em}}
\newcommand{\MSbar}{\overline{\text{MS}}}
\newcommand{\N}{NLO}

\newcommand{\NR}{NLO$\times$RGR}
\newcommand{\NLR}{(NLO+LRR)$\times$RGR}
\newcommand{\NN}{NNLO}

\newcommand{\NNR}{NNLO$\times$RGR}
\newcommand{\NNLR}{(NNLO+LRR)$\times$RGR}

\begin{document}
\title{Pion Valence Quark Distribution at Physical Pion mass of \texorpdfstring{\boldmath{$N_f=2+1+1$}}{Nf=2+1+1} Lattice QCD}

\author{Jack Holligan}
\email{holligan@msu.edu}
\author{Huey-Wen Lin}
\email{hwlin@pa.msu.edu}
\affiliation{Department of Physics and Astronomy, Michigan State University, East Lansing, MI 48824}

\preprint{MSUHEP-23-032}


\begin{abstract}
We present a state-of-the-art calculation of the unpolarized pion valence-quark distribution in the framework of large-momentum effective theory (LaMET)
with improved handling of systematic errors as well as two-loop perturbative matching. 
We use lattice ensembles generated by the MILC collaboration at lattice spacing $a\approx 0.09$~fm, lattice volume $64^3\times 96$, $N_f=2+1+1$ flavors of highly-improved staggered quarks and a physical pion mass. 
The LaMET matrix elements are calculated with pions boosted to momentum $P_z\approx 1.72$~GeV with high-statistics of $O(10^6)$ measurements. 
We study the pion PDF in both hybrid-ratio and hybrid-regularization-independent momentum subtraction (hybrid-RI/MOM) schemes and also compare the systematic errors with and without the addition of leading-renormalon resummation (LRR) and renormalization-group resummation (RGR) in both the renormalization and lightcone matching.
The final lightcone PDF results are presented in the modified minimal-subtraction scheme at renormalization scale $\mu=2.0$~GeV. 
We show that the $x$-dependent PDFs are compatible between the hybrid-ratio and hybrid-RI/MOM renormalization with the same improvements. 
We also show that systematics are greatly reduced by the simultaneous inclusion of RGR and LRR and that these methods are necessary if improved precision is to be reached with higher-order terms in renormalization and matching. 
\end{abstract}

\maketitle

\section{Introduction}\label{sec.Introduction}

Parton distribution functions (PDFs) were first introduced by Feynman in 1969~\cite{Feynman:1969wa} and describe the distribution of longitudinal momentum among a hadron's constituent quarks and gluons.
They are of particular interest due to their use as inputs for the computation of scattering cross sections in high-energy hadron collisions, since the sum of convolutions of PDFs with the cross section of the corresponding parton produces a leading-order approximation of the hadronic cross section in the collinear framework~\cite{Martin:2009iq}.
In addition, they provide insight into the internal structure of the corresponding hadron.
A great deal has been learned about nucleon PDFs from analysis of hard-scattering experiments since the 1960s (for reviews and latest results, see Refs.~\cite{Amoroso:2022eow,Ethier:2020way,Lin:2017snn,Constantinou:2020hdm,Lin:2023kxn}), and these measurements have provided a standard against which theoretical calculations can be judged.
Still greater experimental precision is anticipated at the future Electron-Ion Collider~\cite{AbdulKhalek:2021gbh,Accardi:2012qut,Burkert:2022hjz}.
The pion is the lightest hadron and is of special interest due to its interpretation as a pseudo--Nambu-Goldstone boson from the spontaneous breaking of approximate chiral symmetry;
hence, an understanding of the pion PDF is of high value.

PDFs can be determined numerically (as well as from experimental data) in lattice quantum chromodynamics (QCD) with the current-current correlator method~\cite{Braun:2007wv,Ma:2017pxb}, the pseudo-PDF method~\cite{Radyushkin:2017cyf,Karpie:2017bzm,Orginos:2017kos} or large-momentum effective theory (LaMET)~\cite{Ji:2013dva,Ji:2014gla,Ji:2020ect}, the third of which we use in this work.
LaMET examines the behavior of equal-time, spatially separated correlators on a Euclidean lattice and recovers the lightcone physics through perturbative matching.
A great deal has been achieved applying the LaMET method to 
calculations of PDFs~\cite{Gao:2021dbh,Gao:2022uhg,Alexandrou:2018pbm,Alexandrou:2019lfo,Lin:2018pvv,Lin:2017ani,Chen:2018xof,LatticeParton:2018gjr,Lin:2019ocg,Fan:2020nzz,Alexandrou:2020qtt,Lin:2020fsj,Lin:2020rxa,Chen:2019lcm,Lin:2021brq,Alexandrou:2020zbe,
Lin:2014zya,Chen:2016utp,Alexandrou:2015rja,Alexandrou:2016jqi,Alexandrou:2017huk,Chen:2017mzz,Alexandrou:2018yuy,Zhang:2018nsy,Alexandrou:2018eet,Fan:2018dxu,Liu:2018hxv}.

The $x$-dependent pion valence-quark PDF has been studied on the lattice in recent years~\cite{Zhang:2018nsy,Lin:2020ssv,Sufian:2020vzb,Joo:2019bzr,Izubuchi:2019lyk,Sufian:2019bol,Gao:2021hxl}, and the field has matured to the point where it is important to control sources of systematic errors.
Improvements have been made in the development of the hybrid renormalization scheme~\cite{Ji:2020brr} (compared to the pure ratio- or pure regularization-independent momentum-subtraction (pure RI/MOM)-schemes), lightcone matching up to two-loop order and the inclusion of Wilson-line extrapolation in the renormalized matrix elements which reduces the presence of unphysical oscillations in the PDF.
Two recent developments which we examine in this paper are the presence of large logarithms and the renormalon ambiguity both in the renormalization and matching processes.
The large logarithms are handled with renormalization group resummation (RGR) \cite{Su:2022fiu} and the renormalon ambiguity is handled with of leading renormalon resummation (LRR) \cite{Zhang:2023bxs}.
The RGR method accounts for the fact that lightcone matching depends on the longitudinal momentum of the parton as well as the final renormalization scale, and the difference between them results in large logarithms, which require resummation.
The RGR technique, as applied to the matching process, chooses the renormalization scale such that the logarithms vanish, and the result is evolved to the final desired scale with the DGLAP (Dokshitzer-Gribov-Lipatov-Altarelli-Parisi) equation.
The LRR method is applied to the renormalization of the bare matrix elements by demanding that the short-distance behavior agrees with the theoretical predictions of the operator product expansion (OPE).
These are known as Wilson coefficients and are computed as a perturbation series in the strong coupling;
however, the series is not convergent due to the presence of an infrared renormalon (IRR).
The LRR method resums the terms due to the IRR and improves the convergence of the perturbative calculation.
The two methods of RGR and LRR in both the renormalization and matching processes performed up to next-to-next-to-leading-order (NNLO) result in greatly reduced systematic uncertainties in the computed PDF.
The method of RGR in the matching process has been applied to the pion PDF~\cite{Su:2022fiu} as well as the nucleon transversity PDF~\cite{LatticeParton:2022xsd}. 
The RGR and LRR methods were applied simultaneously to the pion PDF in Ref.~\cite{Zhang:2023bxs} for the renormalization and lightcone matching processes; only the systematic errors were presented.

In this paper we apply both RGR and LRR methods on top of different renormalization schemes to examine their effects on the pion valence-quark PDFs and their systematic errors.
We use clover fermions at a lattice spacing of $a\approx 0.09$~fm and box length $L=64a \approx 5.76$~fm with $N_f=2+1+1$ flavors of highly-improved staggered quarks. The parameters are tuned so as to produce a pion of mass $m_{\pi}\approx 130$~MeV~\cite{Lin:2023gxz}.
The lattice configurations were generated by the MILC collaboration~\cite{MILC:2012znn,MILC:2010pul,MILC:2015tqx}.
The pion matrix elements is calculated with a boosted momentum of $P_z=8\times\frac{2\pi}{L}\approx 1.72$~GeV with the number of measurements for each source-sink separation up to $O(10^6)$. However, Ref.~\cite{Lin:2023gxz}
only reports pion valence-quark PDF using the hybrid-ratio renormalization scheme with NNLO matching process;
the large logarithms and the renormalon ambiguity were not included in the systemics error estimation.

This paper is organized as follows: in Sec.~\ref{sec.LaMETMethod} we outline the methods of RGR and LRR applied both to the renormalization method 
and the lightcone matching. 
In Sec.~\ref{sec.Results} we display and discuss our results of the renormalized matrix elements and the $x$-dependent PDF, and compare with previous results in the literature.
We conclude our paper in Sec.~\ref{sec.Conclusion}.

\section{Methodology}
\label{sec.LaMETMethod}

In this section we outline the method of hybrid renormalization~\cite{Ji:2020brr} with the additions of RGR and LRR described in Ref.~\cite{Zhang:2023bxs} and with particular emphasis on the linear divergence and renormalon ambiguity.
We then summarize the lightcone matching with both RGR and LRR improvements~\cite{Zhang:2023bxs,Su:2022fiu} which we will be used in Sec.~\ref{sec.Results}.

\subsection{Improved Hybrid Renormalization Scheme}\label{sec.Renormalization}

The matrix elements used in LaMET require renormalization to remove both ultraviolet (UV) and infrared (IR) divergences.
The bare matrix elements used in LaMET are spatially separated correlators
\begin{equation}\label{eq.hB}
  h^\text{B}_{\pi}(z,P_z) = \mel**{\pi(P_z)}{\overline{\psi}(z)\gamma_tW(z,0)\psi(0)}{\pi(P_z)},
\end{equation}
where $\psi$, $\overline{\psi}$ are the bare quark field,
$\ket{\pi(P_z)}$ is the pion state with boost momentum $P_z$ in the $z$ direction, $\gamma_t$ is a Dirac matrix and $W(z,0)=\hat{P}\exp(-ig\int^{z}_{0}\diff{}{z'}A_z(z'))$ is the path-ordered Wilson line with gluon field $A_{\mu}(z)$ connecting the two spatially separated coordinates $(0,0,0,0)$ and $(0,0,0,z)$.

Early renormalization methods for LaMET matrix elements used in lattice parton calculations were the regularization-independent momentum-subtraction (RI/MOM) scheme~\cite{Martinelli:1994ty} in Refs.~\cite{Zhang:2018nsy,Zhang:2020gaj,Constantinou:2017sej,Stewart:2017tvs,Alexandrou:2017huk,Lin:2018pvv,Chen:2018xof,LatticeParton:2018gjr,Lin:2019ocg,
Lin:2014zya,Chen:2016utp,Alexandrou:2015rja,Alexandrou:2016jqi,Chen:2017mzz,Alexandrou:2018yuy,Alexandrou:2018eet,Fan:2018dxu,Liu:2018hxv,Alexandrou:2018pbm,Alexandrou:2019lfo,Fan:2020nzz} 
and the ratio scheme similar to those used in the pseudo-PDF methods~\cite{Radyushkin:2017cyf,Orginos:2017kos}. The ratio scheme renormalizes the non-zero momentum matrix elements by dividing by them by the equivalent zero-momentum matrix element. The RI/MOM scheme divides the non-zero momentum matrix elements by the tree-level matrix element at a given momentum and energy scale.
Later, improved renormalized schemes, such as hybrid-ratio and hybrid-RI/MOM schemes~\cite{Ji:2020brr}, were proposed to use the ratio and RI/MOM methods, respectively, up to a distance $z_s\approx 0.3$~fm~\cite{Ji:2020brr,Gao:2022uhg,Gao:2021dbh,LatticeParton:2022xsd}, and for large distance, $z>z_s$, the bare matrix elements are instead multiplied by an exponential term designed to remove both the linear divergence and the renormalon ambiguity.
In such a scheme, the hybrid renormalized matrix elements $h^\text{R}_{\pi}(z,P_z)$ are given by
\begin{equation}\label{eq:hR}
  h^\text{R}_{\pi}(z,P_z) = \begin{cases}
    N \frac{h^\text{B}_{\pi}(z,P_z)}{Z(z)} &\text{if } z < z_s \\
    N e^{(\delta m+m_0)(z-z_s)} \frac{h^\text{B}_{\pi}(z,P_z)}{Z(z_s)} &\text{if } z \geqslant z_s
  \end{cases},
\end{equation}
where $Z(z)$ can be the bare matrix element at zero momentum $h^\text{B}_{\pi}(z,P_z=0)$ for the hybrid-ratio or RI/MOM factor or $Z^\text{RI/MOM}(z,\mu_\text{RI}, p_R^z=0)$ for the hybrid-RI/MOM scheme;
$\delta m$ and $m_0$ are the linear divergence and renormalon ambiguity, respectively;
the normalization $N = Z(0)/h^\text{B}_{\pi}(z=0,P_z)$ sets the matrix element $h^\text{R}_{\pi}(0,P_z)=1$, satisfying conservation of charge.
The constants $Z(z_s)$ and $e^{-(\delta m+m_0)z_s}$ enforce continuity at $z=z_s$.

The linear divergence arises from the self-energy of the Wilson line~\cite{Ji:2020brr,LatticePartonCollaborationLPC:2021xdx} and can be determined by fitting $Z(z) = B e^{-\delta m\, z}$, whether it is in the hybrid-ratio or hybrid-RI/MOM scheme.
The determination of the linear divergence would appear to be a source of systematic errors;
however, the renormalon ambiguity conspires to cancel this error such that the sum $\delta m+m_0$ that appears in Eq.~\eqref{eq:hR} is fixed.
To determine the renormalon ambiguity, we demand that the renormalized matrix elements agree with the OPE at short distances, $z\lesssim 0.2$~fm.
These are functions of the Wilson coefficients $C_0(z,\mu)$, which can be computed as a perturbation series in the strong coupling.
However, such series are, in general, not convergent to all orders.
The asymptotic nature of the perturbation series results in an uncertainty known as the ``renormalon ambiguity''~\cite{Fischer:1999qm}. 

Having determined the linear divergence, we determine the renormalon ambiguity by fitting the bare matrix elements to
\begin{equation}\label{eq.m0formula}
  e^{(\delta m+m_0)z}h^\text{B}_{\pi}(z,P_z)
  =C_0(z,\mu) + \mathcal{O}\left(z^2\LambdaQCD^2\right)
\end{equation}
at short distances as in Ref.~\cite{Zhang:2023bxs}. 
The unpolarized Wilson coefficient in Eq.~\eqref{eq.m0formula} has been computed up to NNLO~\cite{Li:2020xml,Izubuchi:2018srq} and can be improved with one or both of RGR and LRR~\cite{Zhang:2023bxs}.
The different schemes result in different central values and uncertainties for the renormalon ambiguity.
The unpolarized Wilson coefficients with Wilson length $z$ calculated in the modified minimal-subtraction ($\MSbar$) scheme at scale $\mu$, are
\begin{equation}
    C_0^\text{NLO}(z,\mu) =1+\frac{\alpha_s(\mu)C_F}{2\pi}\left(\frac{3}{2}l(z,\mu)+\frac{5}{2}\right)\label{eq.CNLO}
\end{equation}
at NLO~\cite{Izubuchi:2018srq} and
\begin{multline}
    C_0^\text{NNLO}(z,\mu) = C_0^\text{NLO}(z,\mu) + \left(\frac{\alpha_s(\mu)}{2\pi}\right)^2 \\
    \times \Bigg[l^2(z,\mu) \left(\frac{15}{2} - \frac{n_f}{3}\right)
    + l(z,\mu) \left(37.1731-\frac{5}{3}n_f\right) \\
    - 4.34259 n_f + 51.836 \Bigg]
\end{multline}
at NNLO~\cite{Li:2020xml},
where $l(z,\mu) = \ln\left(z^2\mu^2e^{2\gamma_E}/4\right)$, $\gamma_E$ is the Euler-Mascheroni constant,
$\alpha_s(\mu)$ is the strong coupling at energy scale $\mu$,
$C_F$ is the quadratic Casimir for the fundamental representation of SU(3)
and $n_f$ is the number of fermion flavors.

The Wilson coefficients depend on the renormalization scale $\mu$, as well as the intrinsic physical scale, and the difference between them results in logarithmic terms that need to be resummed.
We perform the resummation using the renormalization-group equation (RGE)
\begin{equation}\label{eq.RGE}
    \frac{\diff{}{C_0(z,\mu)}}{\diff{}{\ln(\mu^2)}}=\gamma(\mu)C_0(z,\mu)
\end{equation}
where $\gamma(\mu)$ is the anomalous dimension, which has been calculated up to three loops~\cite{Braun:2020ymy}.
We can set the energy scale such that the logarithms vanish and then evolve the Wilson coefficient to the desired energy scale by solving Eq.~\eqref{eq.RGE}:
\begin{align}\label{eq.CNLORGR}
    C_0^\text{(N)NLO$\times$RGR}(z,\mu)&=C_0^\text{(N)NLO}(z,\mathtt{z}^{-1}) 
    \times  \mathcal{I}(\mu,\mathtt{z}^{-1}),
\end{align}
where $\mathtt{z}^{-1}\equiv 2e^{-\gamma_E}/z$,
$\mathcal{I}(\mu,\mathtt{z}^{-1})$ is defined by
\begin{equation}
\mathcal{I}(\mu,\mathtt{z}^{-1}) = \exp\left(\int^{\alpha_s(\mu)}_{\alpha_s(\mathtt{z}^{-1})}\diff{}{\alpha'} \frac{\gamma(\alpha')}{\beta(\alpha')}\right),
\end{equation}
$\beta(\alpha_s)$ is the QCD $\beta$ function,
and $\gamma$ must be computed to the same order as $C_0(z,\mathtt{z}^{-1})$.

The Wilson coefficients themselves are determined by a perturbation series in the strong coupling which, in general, is not convergent to all orders resulting in the renormalon ambiguity.
We can account for this by improving the Wilson coefficients with LRR. 
Reference~\cite{Zhang:2023bxs} (motivated by Refs.~\cite{Bali:2013pla,Pineda:2001zq}) suggests modifying the Wilson coefficient according to
\begin{align}\label{eq.CNLOLRR}
    C^{\text{N}^{k}\text{LO+LRR}}_0(z,\mu)&=C^{\text{N}^{k}\text{LO}}_0(z,\mu)\nonumber\\
    &+z\mu\left(C_\text{PV}(z,\mu)-\sum_{i=0}^{k-1}\alpha^{i+1}_s(\mu)r_i\right)
\end{align}
where $k=1$ for NLO, $k=2$ for NNLO,
and
\begin{align}\label{eq.BorelInt}
    C_\text{PV}(z,\mu)&=N_m\frac{4\pi}{\beta_0}\int^{\infty}_{0,\rm PV}\diff{}{u}\exp(-\frac{4\pi u}{\alpha_s(\mu)\beta_0})\nonumber\\
    \times\frac{1}{(1-2u)^{b+1}}&\left(1+c_1(1-2u)+c_2(1-2u)^2+\dots\right).
\end{align}
The subscript ``PV" denotes the Cauchy principal value to regulate the poles in the integrand, and the various parameters are
\begin{align}
    b  ={}& \frac{\beta_1}{2\beta_0}\nonumber\\
    c_1 ={}& \frac{1}{4b}\frac{\beta_1^2-\beta_0\beta_2}{\beta_0^4}\nonumber\\
    c_2 ={}& \frac{1}{32\beta_0^8b(b-1)}\left(\beta_1^4+4\beta_0^3\beta_1\beta_2-2\beta_0\beta_1^2\beta_2\right.\nonumber\\
    &{}+\left.\beta_0^2(-2\beta_1^3+\beta_2^2-2\beta_0^4\beta_3)\right)\nonumber\\
    r_n ={}& N_m\left(\frac{\beta_0}{2\pi}\right)^n\frac{\Gamma(1+n+b)}{\Gamma(1+b)}\nonumber\\
    &{}\times\left(1+\frac{c_1b}{b+n}+\frac{c_2b(b-1)}{(n+b)(n+b-1)}\right)\nonumber\\
    N_m ={}& \begin{cases}
        0.5749687262865643\quad\text{for }n_f=3\\
        0.5522713118193284\quad\text{for }n_f=4\\
        0.5235323457364502\quad\text{for }n_f=5\\
    \end{cases},\nonumber
\end{align}
where $\beta_n$ is the $n^\text{th}$ coefficient of the QCD beta function, and $\Gamma(n)$ is the Euler Gamma function.
The expression in Eq.~\eqref{eq.BorelInt} is applicable to all spin states at twist-two. 
We can combine the two methods of RGR and LRR to make the final high-quality Wilson coefficient
\begin{multline}\label{eq.CNLORGRLRR}
    C_0^\text{((N)NLO+LRR)$\times$RGR}(z,\mu) = C_0^\text{(N)NLO+LRR}(z,\mathtt{z}^{-1}) \\
    \times\mathcal{I}(\mu,\mathtt{z}^{-1}).
\end{multline}
Having improved the Wilson coefficients with RGR and LRR, we can determine the renormalon ambiguity $m_0$ using Eq.~\eqref{eq.m0formula} and fully renormalize the matrix elements in the hybrid scheme using Eq.~\eqref{eq:hR}.

The lightcone PDF $q^\text{v}_{\pi}(x,\mu)$ is obtained by applying the perturbative matching to the quasi-PDF, $\tilde{q}^\text{v}_{\pi}(x,P_z)$ 
which is related to the coordinate space matrix elements via a Fourier transform:
\begin{equation}\label{eq.FT}
    \tilde{q}^\text{v}_{\pi}(x,P_z)=\int^{\infty}_{-\infty}\frac{P_z\diff{}{z}}{\pi}e^{ixzP_z}h^\text{R}_{\pi}(z,P_z).
\end{equation}
The variable $x$ is Bjorken $x$, the fraction of the hadron's longitudinal momentum carried by the valence quark. 
In order to prevent unphysical oscillations in the quasi-PDF, we first extrapolate the renormalized matrix element $h^\text{R}(z,P_z)$ to infinite distance before Fourier transforming using the model proposed in Refs.~\cite{Gao:2021dbh,Gao:2022uhg}:
\begin{equation}\label{eq.Extrapolation}
    h^\text{R}(z,P_z)\to \frac{Ae^{-mz}}{|zP_z|^d}\quad\text{as $z\to\infty$}
\end{equation}
where $A$, $m$ and $d$ are fitting parameters.
This extrapolation model is inspired by the PDF having the functional form $q^\text{v}_{\pi}(x,\mu)\sim x^{d-1}$ at small $x$ values, which corresponds to the anticipated large-distance behavior in the renormalized matrix element in the above equation.
This was also applied to the nucleon in Ref.~\cite{Gao:2022uhg}. 
With the renormalized coordinate-space matrix element computed and extrapolated to infinite length of Wilson-line displacement, we can then compute the quasi-PDF with Eq.~\eqref{eq.FT} and follow by performing the lightcone matching.

\subsection{Lightcone matching}\label{sec.Matching}

The matching process aligns the UV behavior of the quasi-PDF with that of the lightcone PDF $q^\text{v}_{\pi}(x,\mu)$.
The two quantities are related through the matching formula
\begin{multline}\label{eq.Matching}
    \tilde{q}^\text{v}_{\pi}(x,P_z) = \int^{1}_{-1}\frac{\diff{}{y}}{|y|}\mathcal{K}(x,y,\mu,P_z,z_s) q^\text{v}_{\pi}(y,\mu) \\
    +\mathcal{O}\left(\frac{\LambdaQCD^2}{P_z^2x^2(1-x)}\right),
\end{multline}
where $\mathcal{K}$ is the matching kernel.
The corrections to the matching process arise from the fact that the quasi-PDF is computed at finite momentum, whereas the lightcone PDF is defined at infinite momentum~\cite{Gao:2021dbh,Braun:2018brg}.
The full matching kernel is
\begin{equation}\label{eq.MatchingFull}
  \mathcal{K}(x,y,\mu,P_z,z_s) = \mathcal{K}^\text{H-ratio, H-RI/MOM} + \Delta\mathcal{K}^\text{LRR},
\end{equation}
depending on whether we renormalize in the hybrid-ratio or hybrid-RI/MOM scheme.
The matching kernel has been computed up to NNLO for unpolarized quasi-PDFs renormalized in the hybrid-ratio scheme with LRR~\cite{Chen:2020ody,Li:2020xml,Su:2022fiu} as well as to NLO in the RI/MOM scheme in Refs.~\cite{Stewart:2017tvs,Chou:2022drv}.\footnote{When the RI/MOM matrix elements are computed at momentum $p_R=0$, the hybrid-ratio and hybrid-RI/MOM matching kernels coincide at NLO as was shown in Ref.~\cite{Chou:2022drv}.}

The LRR modification~\cite{Zhang:2023bxs} in the matching kernel can be written down as
\begin{multline}
\label{eq.MatchingLRR}
  \Delta\mathcal{K}^\text{LRR} = \int\frac{yP_z\diff{}{z}}{2\pi} e^{i(x-y)zP_z} \\
  \times z\mu\left[C_\text{PV}(z,\mu) - \sum_{i=0}^{k -1} \alpha^{i+1}_s(\mu)r_i\right].
\end{multline}
with $k=1$ for NLO and $k=2$ for NNLO.
Since the matching process can be numerically expensive, we convert the matching kernel $\mathcal{K}$ of Eq.~\eqref{eq.MatchingFull} into a matrix in $x$ and $y$, $\mathcal{K}_{xy}$, then multiply a vector of quasi-PDF values by the inverse $\mathcal{K}^{-1}_{xy}$.

The method of RGR can also be applied to the matching process; the algorithm has been derived in Ref.~\cite{Su:2022fiu}.
This time, the intrinsic physical scale is that of the parton ($\sim 2xP_z$). We perform the lightcone matching at the scale $\mu=2xP_z$ so the logarithms vanish, and we then evolve to the desired energy scale using the DGLAP equation:
\begin{equation}\label{eq.DGLAP}
  \frac{\diff{}{q^\text{v}_{\pi}(x,\mu)}}{\diff{}{\ln(\mu^2)}} =
  \int^1_x \frac{\diff{}{z}}{|z|} \mathcal{P}(z) q^\text{v}_{\pi}\left(\frac{x}{z},\mu\right),
\end{equation}
where $\mathcal{P}(z)$ is the DGLAP kernel, which has been calculated up to three 
loops~\cite{Moch:2004pa}. 
It should be noted that the DGLAP evolution formula begins to break down at $x\approx 0.2$ where $\alpha_s(\mu=2xP_z)$ becomes nonperturbative.

\section{Results and Discussion}\label{sec.Results}

\subsection{Renormalization Matrix Elements}\label{sec:RME}

The first stage in the calculation is to renormalize the bare matrix elements according to Eq.~\eqref{eq:hR}. We determine the linear divergence by fitting $Z(z)$ of Eq.~\eqref{eq:hR} to an exponential decay $Be^{-\delta m\, z}$.
Our calculation from examining the exponential decay of $Z(z)$ yields $\delta m=0.713(13)$ and $\delta m=0.668(10)$~GeV with $Z(z)$ inputs from the RI/MOM renormalization factors and the zero boost-momentum bare pion matrix elements respectively.
Our $\delta m$ parameter are of the same order of magnitude as ANL/BNL's~\cite{Gao:2021dbh} which was $\delta m=0.7439(59)$~GeV at $N_f=2+1$ $a=0.04$~fm on 310-MeV ensemble.
We then proceed to determine the renormalon ambiguity from the bare boosted-momentum pion matrix elements ${h^\text{B}_{\pi}(z,P_z)}$ and Wilson coefficients as described in Eq.~\eqref{eq.m0formula}.
The renormalon ambiguity $m_0$ is obtained by fitting to a linear function in $z$ according to
\begin{equation}\label{eq.m0formula2}
    \ln\left(\frac{e^{-\delta m\,z}C_0(z,\mu)}{h^\text{B}_{\pi}(z,P_z)}\right)=m_0z+c
\end{equation}
at the three $z$ values $\{z_c-0.02\text{ fm},\,z_c,\,z_c+0.02\text{ fm}\}$ for different central values $z_c$ within the range of validity of the OPE ($z_c\lesssim 0.2$~fm).
The terms $m_0$ and $c$ are fitting parameters.
The Wilson coefficient $C_0(z,\mu)$ in the formula can be determined to NLO or NNLO and may have the improvements of LRR and/or RGR as detailed in Sec.~\ref{sec.Renormalization}.
The coefficient of $z$ in the linear fit is the renormalon ambiguity $m_0$ and the term $c$ is the $y$-intercept, which is irrelevant for this calculation.

\begin{figure*}
    \centering
    \subfigure{\includegraphics[width=0.45\textwidth]{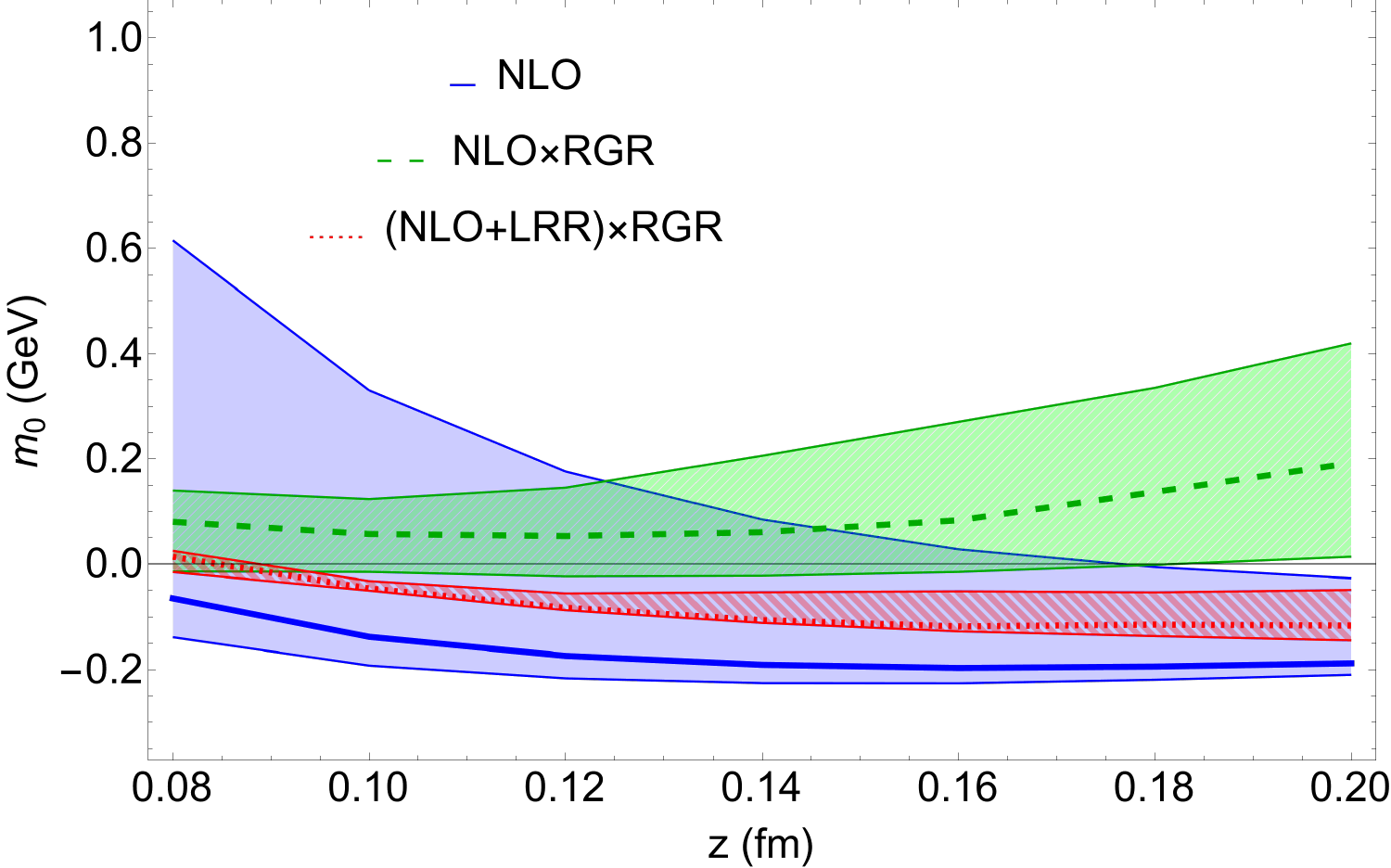}}\quad
    \subfigure{\includegraphics[width=0.45\textwidth]{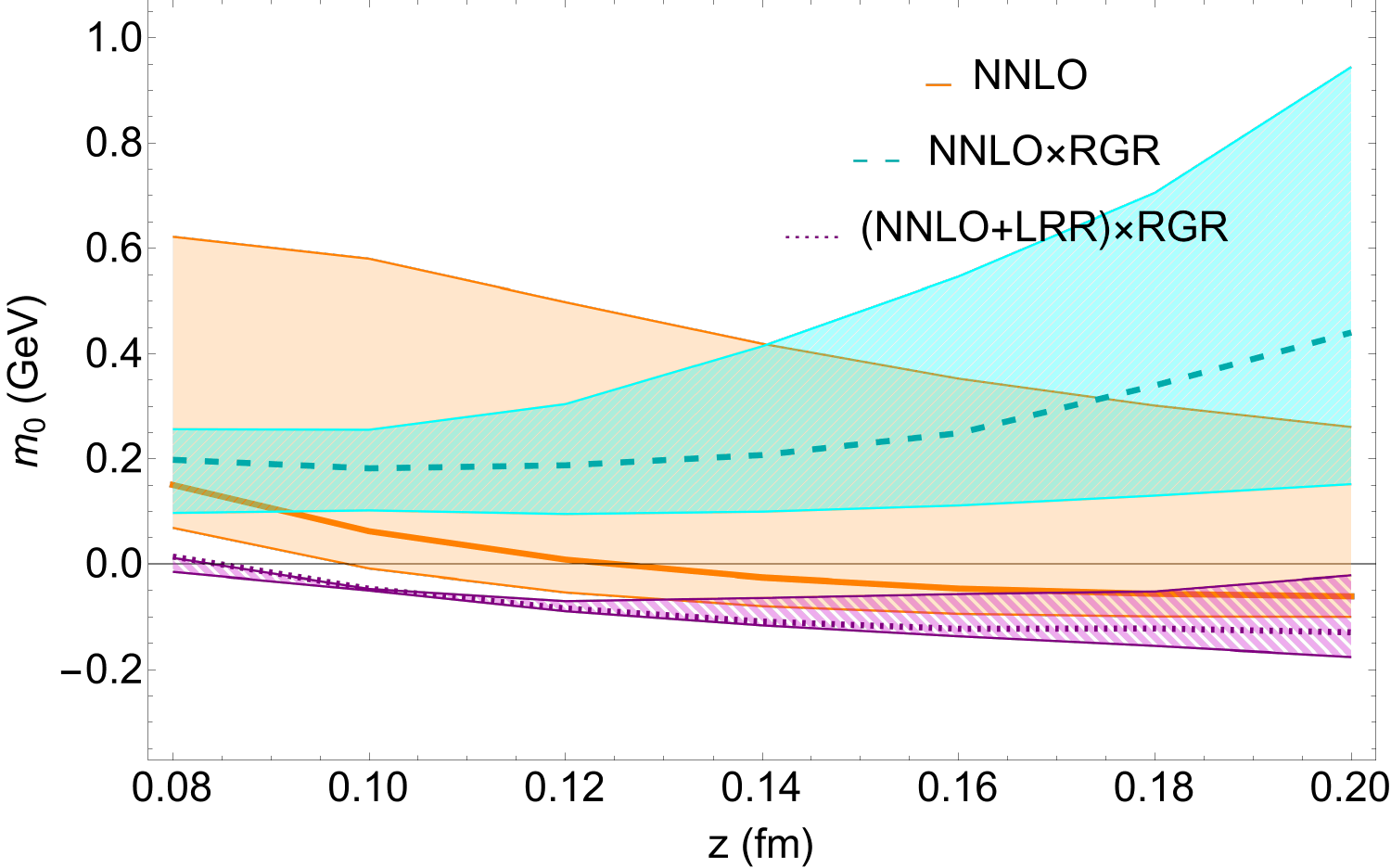}}
    \caption{
    Renormalon ambiguity determined in the hybrid-ratio scheme in multiple $z$ ranges.
    The value $m_0(z)$ is determined by fitting Eq.~(\ref{eq.m0formula2}) to the $z$ values $\{z-0.02\text{ fm},\,z,\,z+0.02\text{ fm}\}$. 
    The vertical width of each band corresponds to the systematic error derived from scale variation described in Sec.~\ref{sec:RME}.
    We show $m_0(z)$ determined at \N, \NR, \NLR~(left figure), \NN, \NNR\ and \NNLR~(right figure) in blue, green, red, orange, cyan and purple, respectively. Note that the errors are minimized in the two cases in which RGR and LRR are applied simultaneously.}
    \label{fig:m0determination}
\end{figure*}

We show the values of the renormalon ambiguity in Fig.~\ref{fig:m0determination} with the Wilson coefficient determined to different orders and in different fitting ranges.
The error bars are derived from ``scale variation".
When we use the RGE, we set the initial scale to $\mu=\mathtt{z}^{-1}$ so as to eliminate the logarithms and then evolve to the final desired energy scale of $\mu=2.0$~GeV.
Scale variation involves setting the initial scale to $c'\times\mathtt{z}^{-1}$ for $c'=0.75$ and $c'=1.5$ as was used in Ref.~\cite{Zhang:2023bxs}.
This corresponds to a variation of approximately $15\%$ either side of $\alpha_s(\mu=2.0\text{ GeV})$ in the strong coupling. 
The central value corresponds to $c'=1.0$.
The numerical results for the renormalon ambiguity with errors derived from scale variation are shown in Tables~\ref{tab:HybridRatioRenorm} and \ref{tab:HybridRIMOMRenorm} for the hybrid-ratio and hybrid-RI/MOM renormalization scheme.
The linear divergence is included in each table for the corresponding scheme.
It can be seen from both the plots in Fig.~\ref{fig:m0determination} as well as Tables~\ref{tab:HybridRatioRenorm} and \ref{tab:HybridRIMOMRenorm} that the calculations with both the LRR and RGR improvements are the most reliable due to their nice plateau behavior in the region $z_c=[0.12, 0.2]$~fm as well as their greatly reduced errors compared to the other schemes.
The improvement of RGR without LRR causes the renormalon ambiguity to become quite unstable and the error bars are very large.
This is an evidence (also present in Ref.~\cite{Zhang:2023bxs}) that the resummation of large logarithms on its own results in the Wilson coefficient being dominated by the renormalon ambiguity.
The renormalon divergence begins to emerge at order $k$; $k\sim 1/\alpha_s(\mu)$.
At an energy scale of $\mu=2.0$~GeV, $1/\alpha_s\approx 3$ and so the renormalon ambiguity will not begin to emerge until N$^3$LO.
However, when RGR is used, the energy scale is set to an initial value of $\mu=2e^{-\gamma_E}/z$.
At short distances, this is a large energy, hence a small coupling and the renormalon will not emerge at order NLO or NNLO.
However, the reverse is the case at large $z$ for which this is a small energy and the strong coupling is large.
The renormalon divergence can be significant even at NLO in this region.
This is the reason that the Wilson coefficients with RGR but not LRR have very large systematic errors at $z\gtrsim 0.2$~fm.
The renormalon divergence must be included when RGR is used for a reliable measurement.

\begin{figure}
    \centering
    \includegraphics[width=0.45\textwidth]{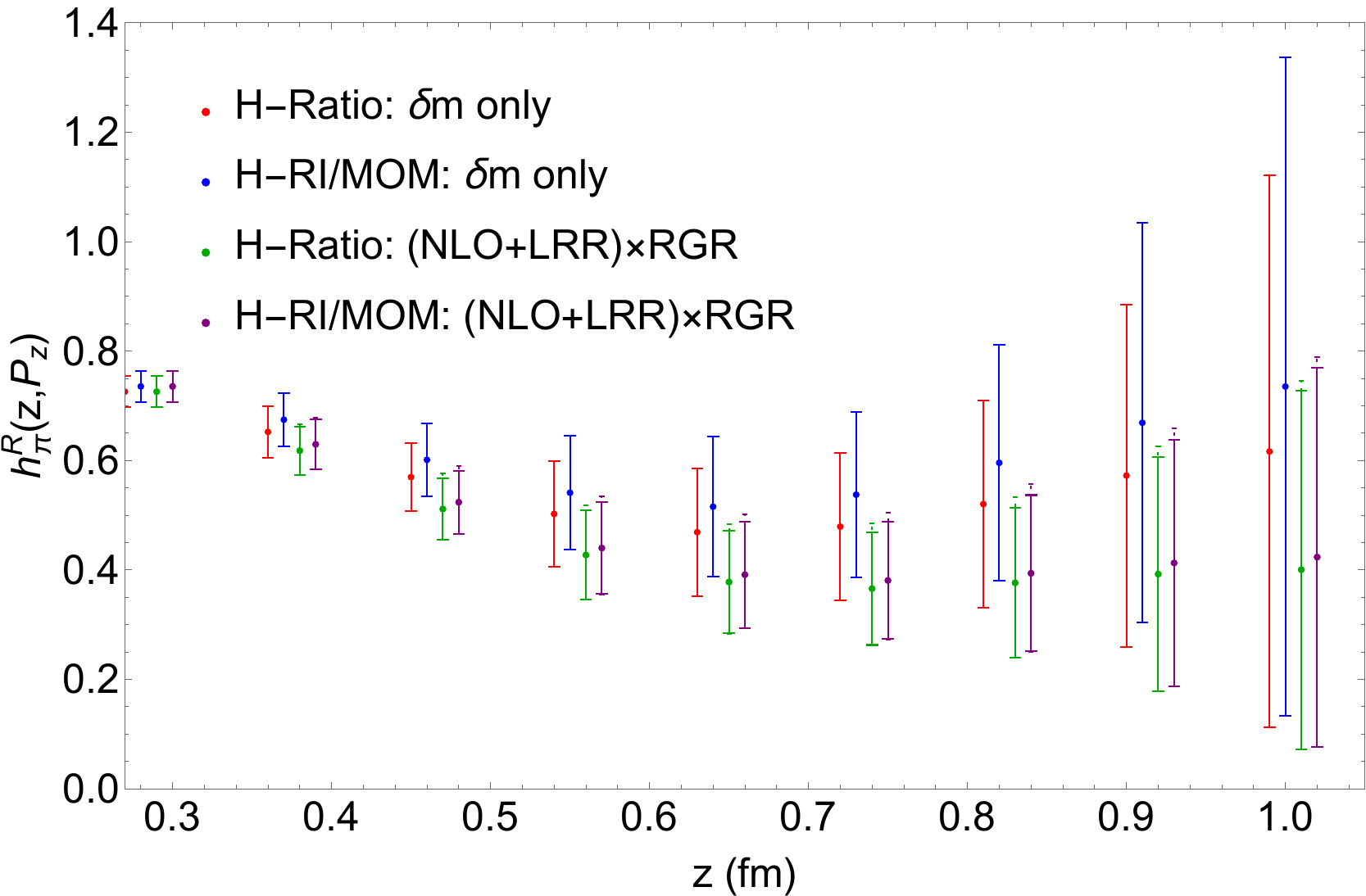}
    \caption{
    $\delta m$ only renormalized matrix elements in the hybrid-ratio (red) and hybrid-RI/MOM (blue) schemes; 
    \NLR\ renormalized matrix elements in the hybrid-ratio (green) and the hybrid-RI/MOM (purple) schemes. In the two cases of \NLR, the solid error bars are statistical and the dashed error bars are combined statistical and systematic, the latter arising from the scale variation.
    Except for the $\delta m$-only calculation in the hybrid-ratio scheme, the plotted data have been offset from their exact $z$ value to allow for readability.}
    \label{fig:ratio-RIMOM}
\end{figure}

The lattice data renormalized in the hybrid-ratio and hybrid-RI/MOM schemes are shown in Fig.~\ref{fig:ratio-RIMOM}.
The plot shows data with the renormalon ambiguity omitted and determined to \NLR\ both in the hybrid-ratio and hybrid-RI/MOM schemes.
We show both statistical errors and, in the cases of \NLR, combined statistical and systematic errors.
It is clear from Fig.~\ref{fig:ratio-RIMOM} that the central values between the hybrid-ratio and hybrid-RI/MOM methods to an otherwise fixed order are compatible within statistical errors.
The central values of the $\delta m$ matrix elements can differ by up to 20\% between the two renormalization schemes but have very large statistical errors.
By contrast, the \NLR\ results differ by no more than 7\% between the two schemes and have much smaller statistical errors.
The systematic errors are also very small in the \NLR\ case.
For this reason, the subsequent steps in the calculation of the pion PDF use the hybrid-ratio scheme since the matching kernel has been computed up to NNLO.

\begin{figure}
    \centering
    \includegraphics[width=0.45\textwidth]{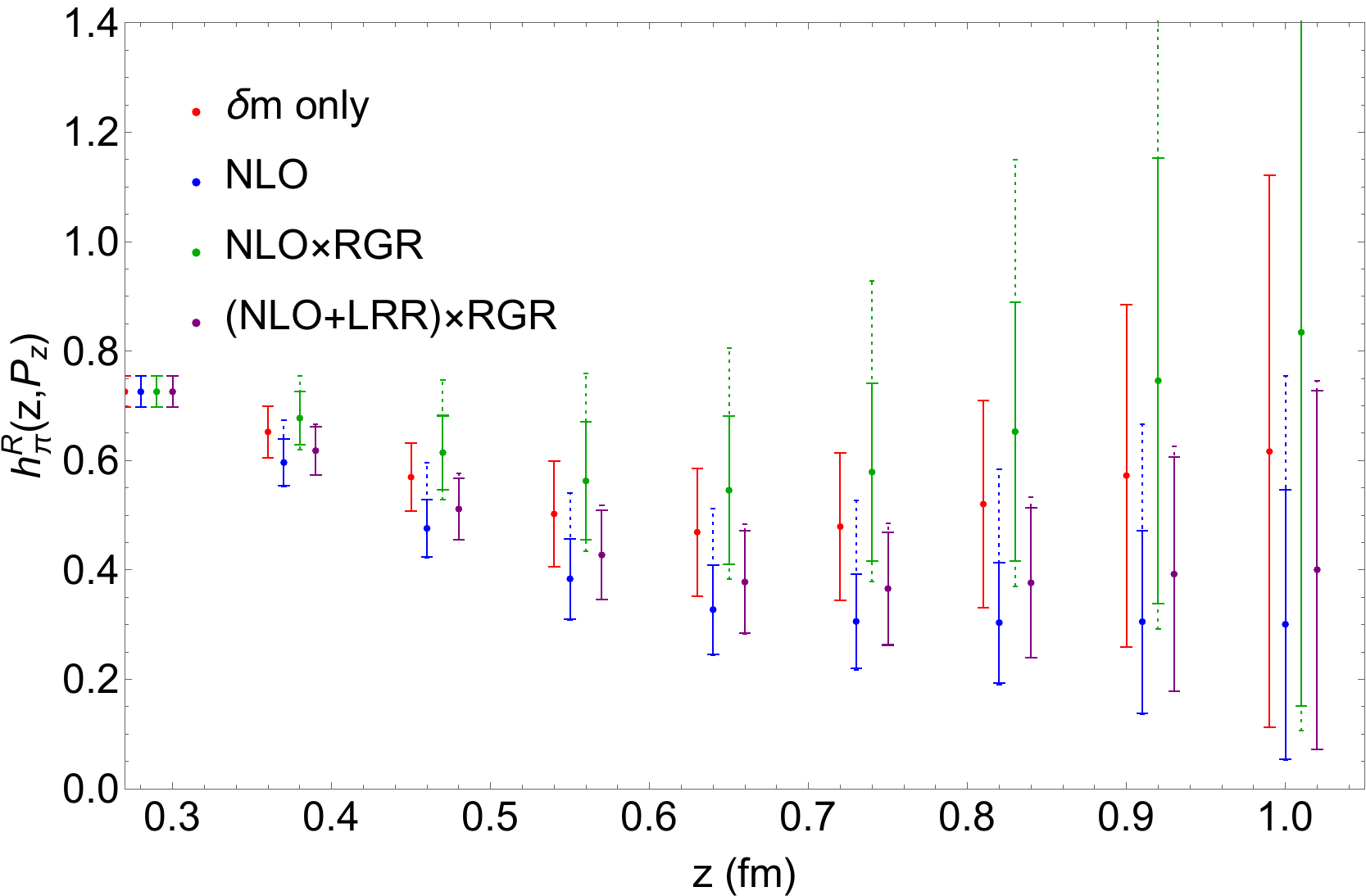}
    \caption{
    $\delta m$-only (red), \N\ (blue), \NR\ (green) and \NLR\ (purple) matrix elements renormalized in the hybrid-ratio scheme. In the cases of \N, \NR\ and \NLR, the solid error bars are statistical and the dashed error bars are combined statistical and systematic errors, the latter arising from the scale variation. The same three sets of plotted data have been offset from their exact $z$ values to allow for readability.}
    \label{fig:hRratio}
\end{figure}

In Fig.~\ref{fig:hRratio}, we show the pion matrix elements renormalized in the hybrid-ratio scheme with $\delta m$ only (red) and with the renormalon ambiguity determined to \N\ (blue), \NR\ (green) and \NLR\ (purple). 
We can see that the systematic errors are minimized when both RGR and LRR improvements are applied simultaneously.
In addition, the statistical and systematic errors are very large when RGR is applied on its own due to the RGR process enhancing the presence of the renormalon ambiguity.
Between the results at \N\ and those at \NLR, the central values are compatible within statistical errors but the latter have much smaller systematic errors compared to the former.
The overall $z$-behavior changes among the four schemes due to $m_0$ in the exponential term that renormalizes the matrix elements for $z>z_s$ in Eq.~(\ref{eq:hR}).
Only \NR\ has a positive $m_0$ value; hence, the corresponding renormalized matrix elements have the largest central values.
By contrast, $m_0$ at \NLR\ is negative and \N\ is even smaller.
Hence, the \NLR\ renormalized matrix elements are smaller than the $\delta m$ terms and the \N\ are smaller still.
A greater suppression in $z$ results in less pronounced oscillations in the Fourier transform (i.e. the quasi-PDF), so the \N\ and \NLR\ $x$-dependent PDFs are more steady.

\begin{figure}
    \centering
    \includegraphics[width=0.45\textwidth]{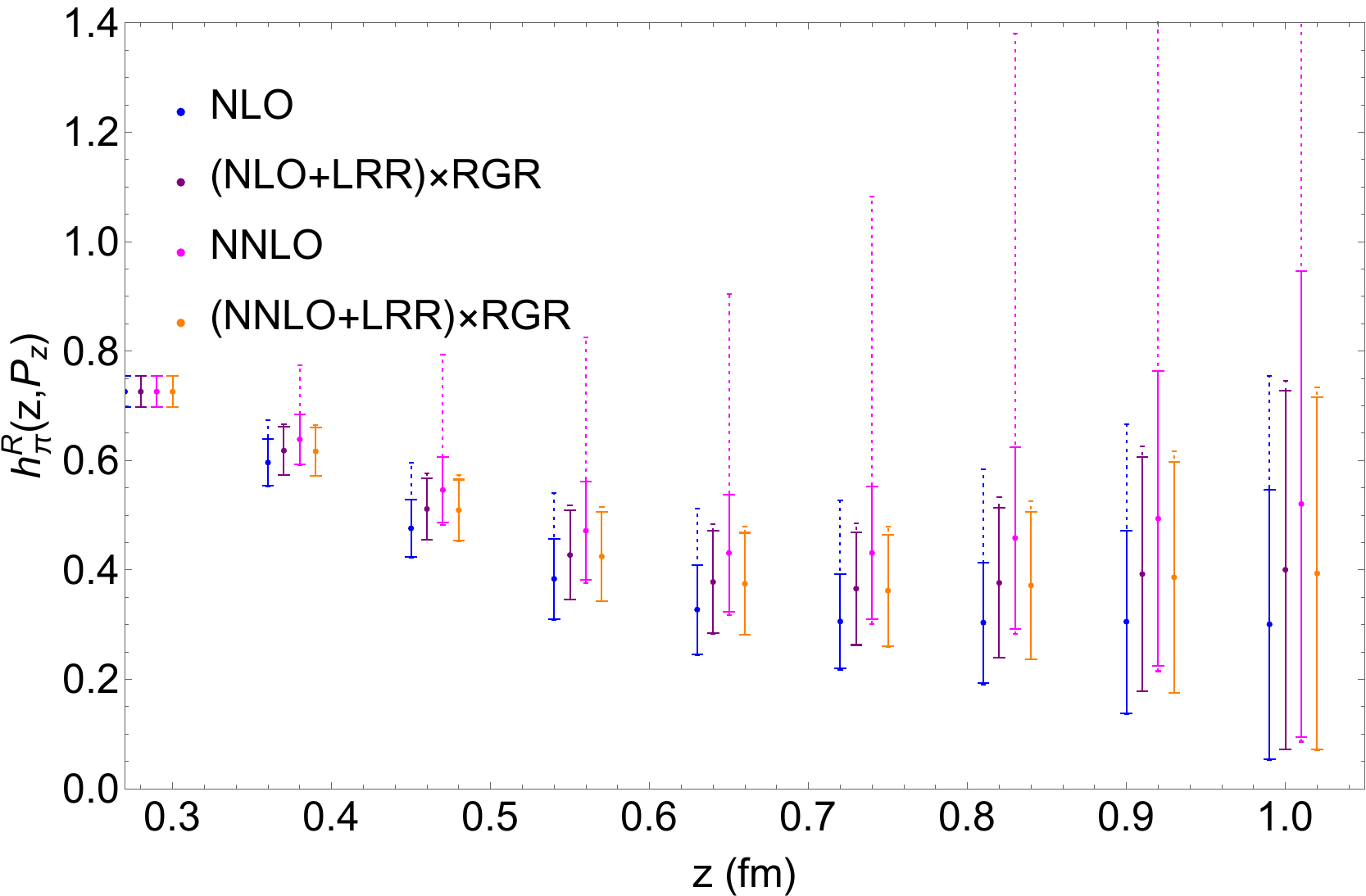}
    \caption{
    \N\ (blue), \NLR\ (purple), \NN\ (magenta) and \NNLR\ (orange) matrix elements renormalized in the hybrid ratio scheme. The solid error bars are statistical and the dashed error bars are combined statistical and systematic errors, the latter arising from the scale variation. All plotted data except for \N\ have been offset from their exact $z$ values to allow for readability.}
    \label{fig:hRNNLO}
\end{figure}

We then compare the renormalized matrix elements at orders \N, \NLR, \NN\ and \NNLR\ in Fig.~\ref{fig:hRNNLO}.
While the lower systematic errors are insignificant with relative to the statistical errors, the upper systematic errors increase from 0.13 to 0.21 at $z=0.36$ fm and from 1.5 to 3.4 at $z=0.99$ fm from \N\ to \NN. 
However, the same errors decrease to 0.08 at $z=0.36$~fm and 0.8 at $z=0.99$~fm from \NN\ to \NNLR.
In addition, the relative systematic errors 
decrease from \NLR\ to \NNLR. 
Not only does this reaffirm the benefits of including RGR and LRR in our calculations, but it also shows that if one wishes to renormalize to order NNLO or higher, it is necessary to account for both large logarithms and the renormalon ambiguity.
In other words, handling of the systematic errors must improve in parallel with higher order terms in the renormalization process.
If systematics are not accounted for in the calculations, the results will deteriorate at higher orders.
This demonstrates the need to account for the large logarithms and the renormalon divergence in the renormalization of LaMET matrix elements.

\begin{table}
    \centering
    \begin{tabular}{|c|c|}
    \hline
    Order & $m_0$ (GeV)\\
    \hline\hline
    $\delta m$ only & 0.0\\
    \hline
    \N & $-0.197^{+0.22}_{-0.029}$\\
    \hline
    \NN & $-0.05^{+0.4}_{-0.05}$\\
    \hline
    \NR & $0.08^{+0.19}_{-0.10}$\\
    \hline
    \NNR & $0.25^{+0.30}_{-0.14}$\\
    \hline
    \NLR & $-0.118^{+0.07}_{-0.009}$ \\
    \hline
    \NNLR & $-0.123^{+0.07}_{-0.014}$\\
    \hline
    \end{tabular}
    \caption{Renormalon ambiguity $m_0$
    for hybrid-ratio scheme determined in the fitting range $z\in [0.14, 0.18]$~fm.
    The errors in $m_0$ are derived from scale variation described in Sec.~\ref{sec:RME}.
    Each $m_0$ is added to the linear divergence $\delta m=0.668(10)$ GeV in the hybrid-ratio scheme for $z>z_s$.
    Note the greatly reduced errors in $m_0$ when RGR and LRR are applied simultaneously.}
    \label{tab:HybridRatioRenorm}
\end{table}

\begin{table}
    \centering
    \begin{tabular}{|c|c|}
    \hline
    Order & $m_0$ (GeV)\\
    \hline\hline
    $\delta m$ only & 0.0\\
    \hline
    NLO & $-0.230^{+0.22}_{-0.029}$\\
    \hline
    NLO$\times$RGR & $0.05^{+0.19}_{-0.10}$\\
    \hline
    (NLO+LRR)$\times$RGR & $-0.151^{+0.07}_{-0.009}$ \\
    \hline
    \end{tabular}
    \caption{Renormalon ambiguity $m_0$ for hybrid-RI/MOM scheme determined in the fitting range $z\in [0.14, 0.18]$~fm.
    The errors in $m_0$ are derived from scale variation described in Sec.~\ref{sec:RME}.
    Each $m_0$ is added to the linear divergence $\delta m=0.713(13)$ GeV in the hybrid-RI/MOM scheme for $z>z_s$.
    Note the greatly reduced errors in $m_0$ when RGR and LRR are applied simultaneously.}
    \label{tab:HybridRIMOMRenorm}
\end{table}

\subsection{Parton Distribution Functions}

To obtain the pion quasi-PDF, we first need to extrapolate the renormalized matrix elements at large Wilson-line displacement to infinity.
The Wilson-line displacement must be large enough ($\gtrsim 0.5$~fm) that we can realistically use the small-$x$ model corresponding to the large distance behavior; 
in this work, we choose the fitting range $[8a, 14a] \approx [0.72, 1.26]$~fm.
Several extrapolation models were tested by the ANL/BNL collaboration
in Ref.~\cite{Gao:2021dbh} for the pion matrix elements and the final one used was that of Eq.~\eqref{eq.Extrapolation} with the constraints $A>0$, $d>0$ and $m>0.1$~GeV; 
the last two constraints ensure that the large distance behavior decays sufficiently quickly to obtain a convergent Fourier transform.  
Once we obtain quasi-PDF, the final step is to apply the matching process, Eq.~(\ref{eq.Matching}), to recover the lightcone PDF.

\begin{figure}
    \centering
    \includegraphics[width=0.45\textwidth]{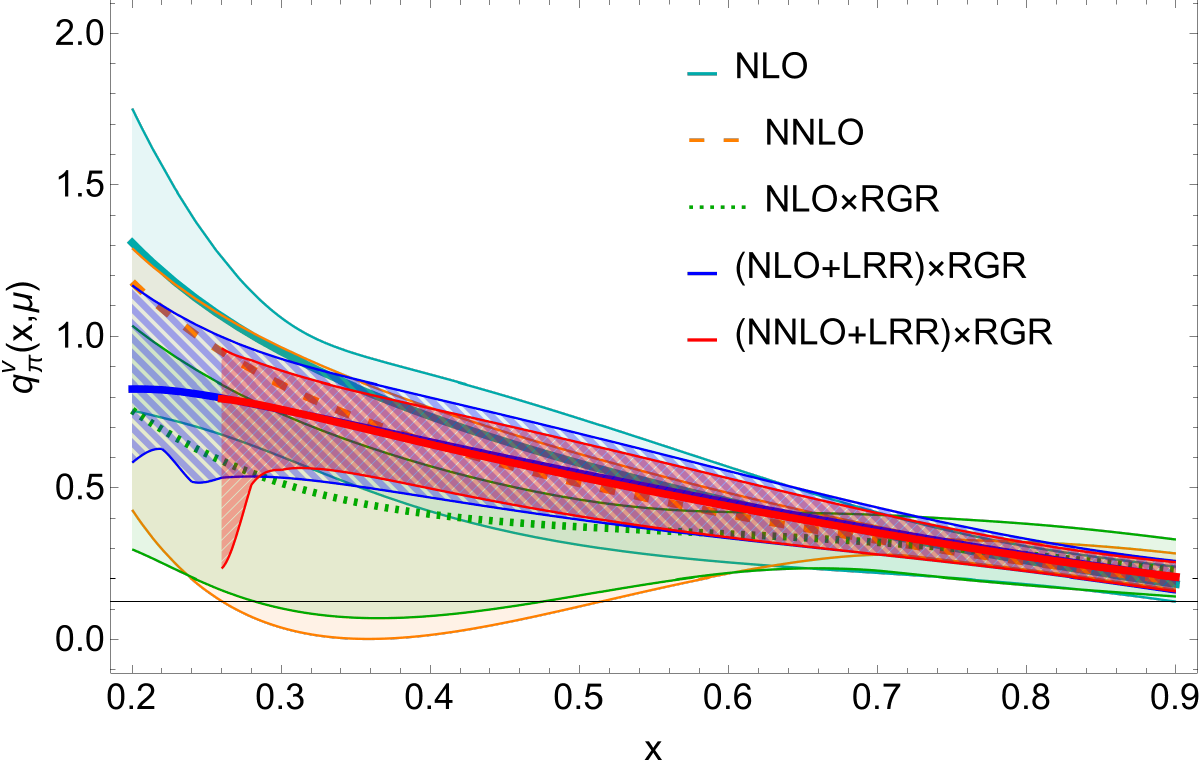}
    \caption{\N\ (cyan), \NN\ (dashed orange), \NR\ (dotted green), \NLR\ (hatched blue) and \NNLR\ (hatched red) systematic errors of the lightcone pion valence-quark PDFs.
    Statistical errors are negligible here.}
    \label{fig:PDFPLB}
\end{figure}

Before we study the full statistical and systematic errors of pion valence-quark PDF, we first examine their $x$-dependent
\N, \NN, \NR, \NLR\ and \NNLR\ systematic errors in Fig.~\ref{fig:PDFPLB}. 
We see that the lower systematic errors are larger for the two orders \NN\ and \NR\ than they are at \N, which was also found by Ref.~\cite{Zhang:2023bxs}.
In addition, the systematic errors decrease significantly from \N, to \NLR, and further reduction from \NLR\ to \NNLR.
We also observe the central values for \NLR\ and \NNLR\ are much closer to each other than those for \N\ to \NN\, showing better convergence going to higher order with LRR and RGR improvements. 
These qualitative behaviors are also consistent with what was found in the earlier pion-PDF study~\cite{Zhang:2023bxs}.

\begin{figure}
    \centering
    \includegraphics[width=0.5\textwidth]{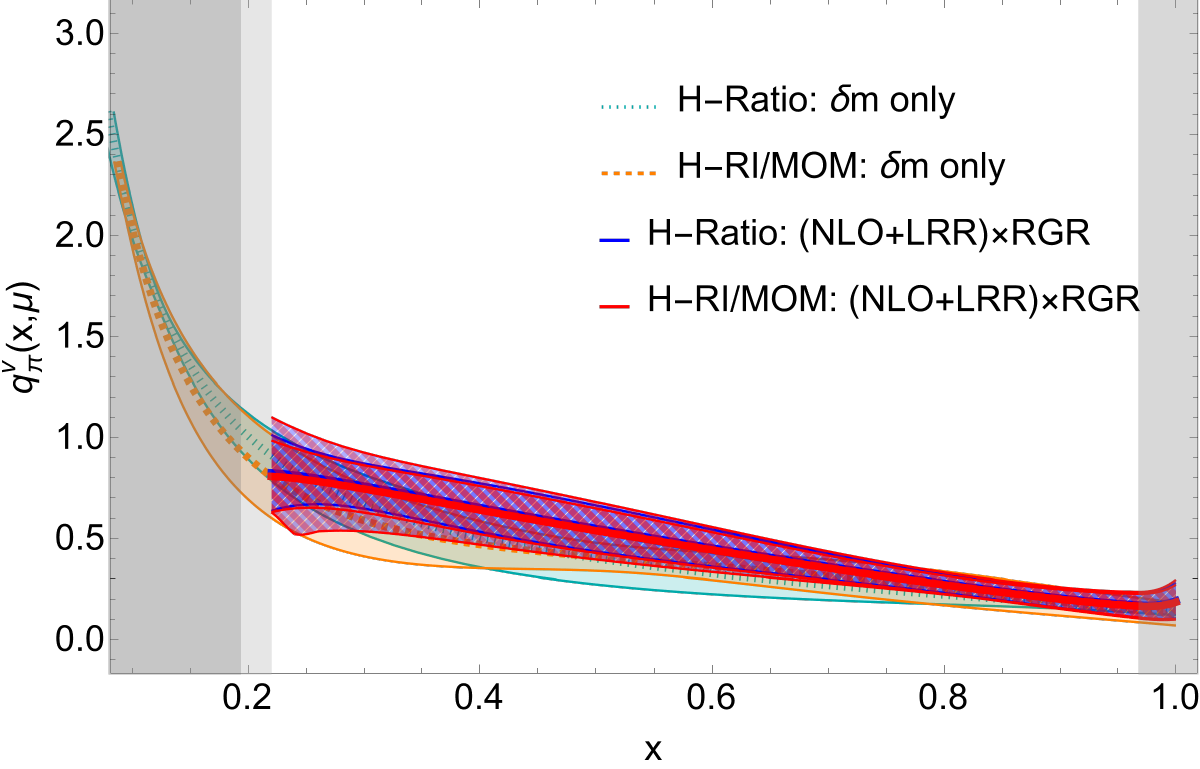}
    \caption{
The lightcone PDFs renormalized in hybrid-ratio and hybrid-RI/MOM schemes
with $\delta m$ only (dotted cyan and dashed orange) and at \NLR~(hatched blue and hatched red).
In each PDF, the inner darker bands show statistical error, while the outer lighter bands show combined statistical and systematic errors (due to scale variation).
The dark-gray region is where the LaMET calculation begins to break down, and the light-gray region is where the RGR matching begins to break down. 
}
    \label{fig:PDF1}
\end{figure}

We compare the pion PDFs from hybrid-RI/MOM and hybrid-ratio in Fig.~\ref{fig:PDF1};
the renormalized matrix elements can be found in Fig.~\ref{fig:ratio-RIMOM}.
The RGR matching process begins to break down at small-$x$, where the strong coupling becomes nonperturbative at energy scale $\mu=2xP_z$.
For this reason, we do not plot the PDF for $x\lesssim 0.22$.
We noted that the two \NLR\ renormalization schemes PDFs produce very similar central values, while the PDFs from the two $\delta m$-only renormalization schemes differ noticeably.
This is understandable, since the PDFs from the $\delta m$-only schemes have different linear-divergence contributions, as discussed in Sec.~\ref{sec:RME}.
Furthermore, the two corresponding \NLR\ renormalized matrix elements have very similar central values, as shown in Fig.~\ref{fig:ratio-RIMOM}.
In addition, the lightcone matching is the same for the two schemes when the RI/MOM matrix elements are evaluated at $p_R^z=0$, as was demonstrated in Ref.~\cite{Chou:2022drv} at NLO, so the matching process does not cause a further differences in the $x$-dependent results for hybrid-ratio and hybrid-RI/MOM schemes.
Overall, we found the pion valence-quark PDF to be consistent between the hybrid-RI/MOM and hybrid-ratio renormalization schemes.
For the rest of this work, we will focus on the pion valence-quark PDF results with hybrid-ratio renormalization scheme.

\begin{figure}
    \centering
    \includegraphics[width=0.5\textwidth]{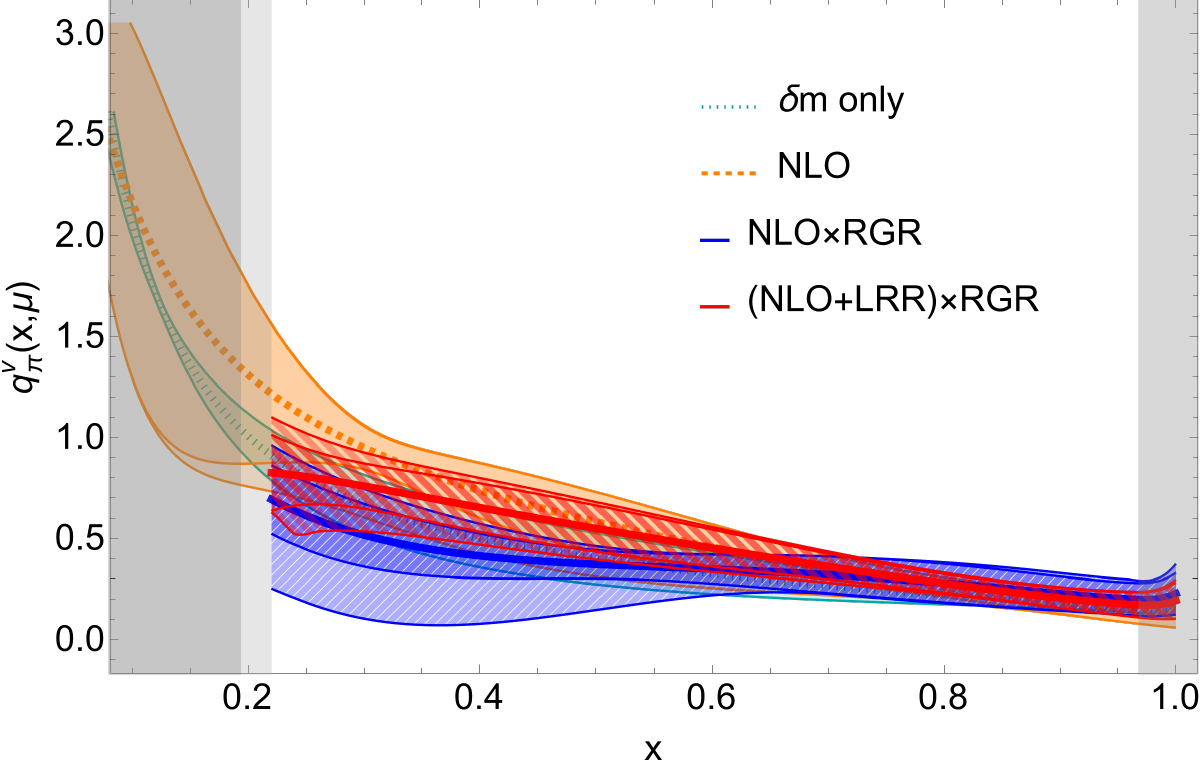}
    \caption{
    The lightcone pion valence-quark PDFs as a function of $x$ renormalized in hybrid-ratio schemes
with  $\delta m$ only (dotted cyan), \N\ (dashed orange), \NR\ (hatched blue) and \NLR\ (hatched red) improvement.
In each PDF, the inner darker bands show statistical error, while the outer lighter bands show combined statistical and systematic errors (due to scale variation).
The dark-gray region is where the LaMET calculation begins to break down, and the light-gray region is where the RGR matching begins to break down.
    }
    \label{fig:PDF2}
\end{figure}

We compare $x$-dependent PDFs in the hybrid-ratio scheme with $\delta m$-only, \N, \NR\ and \NLR\ improvement in Fig.~\ref{fig:PDF2}.
Going from \N\ to \NR, the relative systematic errors increase by as much as 40\%, showing that the RGR process on its own can enhance the presence of the infrared renormalon.
However, the relative systematic errors are reduced by as much as a factor of two when going from \NR\ to \NLR, demonstrating that it is necessary to include the LRR method when RGR is used.
We also see that the central values of the two RGR plots (\NR\ and \NLR) are lower than those of \N\ in the region $x\lesssim 0.6$.
This is to be expected, since the logarithmic terms that are resummed in the RGR matching process are $\ln(\mu^2/4x^2P_z^2)$, which become larger as $x$ decreases.
Thus, their effect is more significant at small $x$, and the effect of resumming them is to lower the central value.

\begin{figure}
    \centering
    \includegraphics[width=0.5\textwidth]{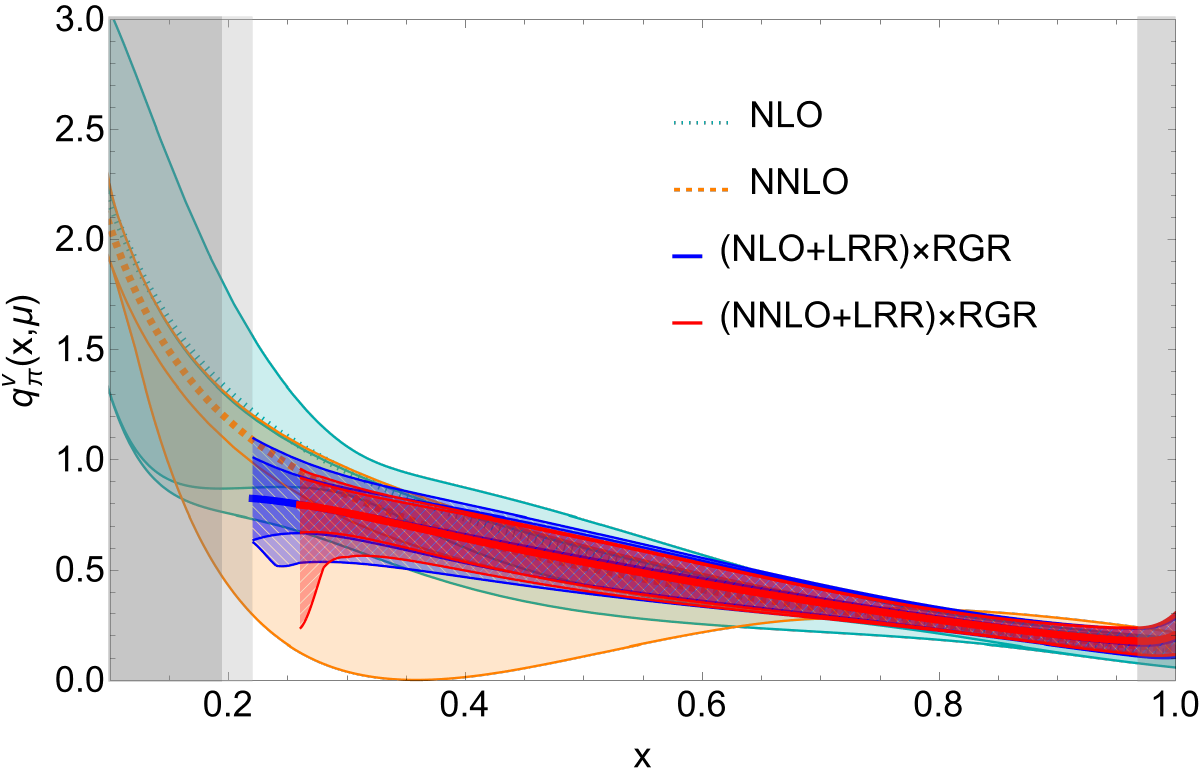}
    \caption{
  The lightcone pion valence-quark PDFs as a function of $x$ renormalized in hybrid-ratio scheme with \N\ (dotted cyan), \NLR\ (hatched blue), \NN\ (dashed orange) and \NNLR\ (hatched red) improvement.
In each PDF, the inner darker band shows statistical error, while the outer lighter band shows combined statistical and systematic errors (due to scale variation).
The dark-gray region is where the LaMET calculation begins to break down, and the light-gray region is where the RGR matching begins to break down.
    }
    \label{fig:PDF3}
\end{figure}

The pion valence-quark PDFs for one- and two-loop treatments are compared in Fig.~\ref{fig:PDF3}:
\N\ (dotted cyan), \NLR\ (solid blue), \NN\ ~(dashed orange) and \NNLR\ (solid red) improvements.
We would normally expect a calculation to yield more precise results as we go to higher order (e.g. \N\ to \NN).
However, this is not the case in Fig.~\ref{fig:PDF3}; the systematic errors increase from \N\ to \NN. 
It is necessary to account for the effects of large logarithms and the renormalon divergence with the methods of RGR and LRR, respectively.
We can see that the systematic errors decrease from \NLR\ to \NNLR, showing that, in this case, the calculation becomes more precise. 
Figure~\ref{fig:PDF3} is a demonstration that it is necessary to control the sources of systematic errors if higher-order terms are to be included in the lightcone matching.
Since the systematic errors decrease from \NLR\ to \NNLR\ by 10\% to 15\%, 
we see the benefits of including higher-order terms in the matching and renormalization.
This demonstrates that it is necessary to include both RGR and LRR if we wish to renormalize and match to two-loop level. 

\begin{figure}
    \centering
    \includegraphics[width=0.5\textwidth]{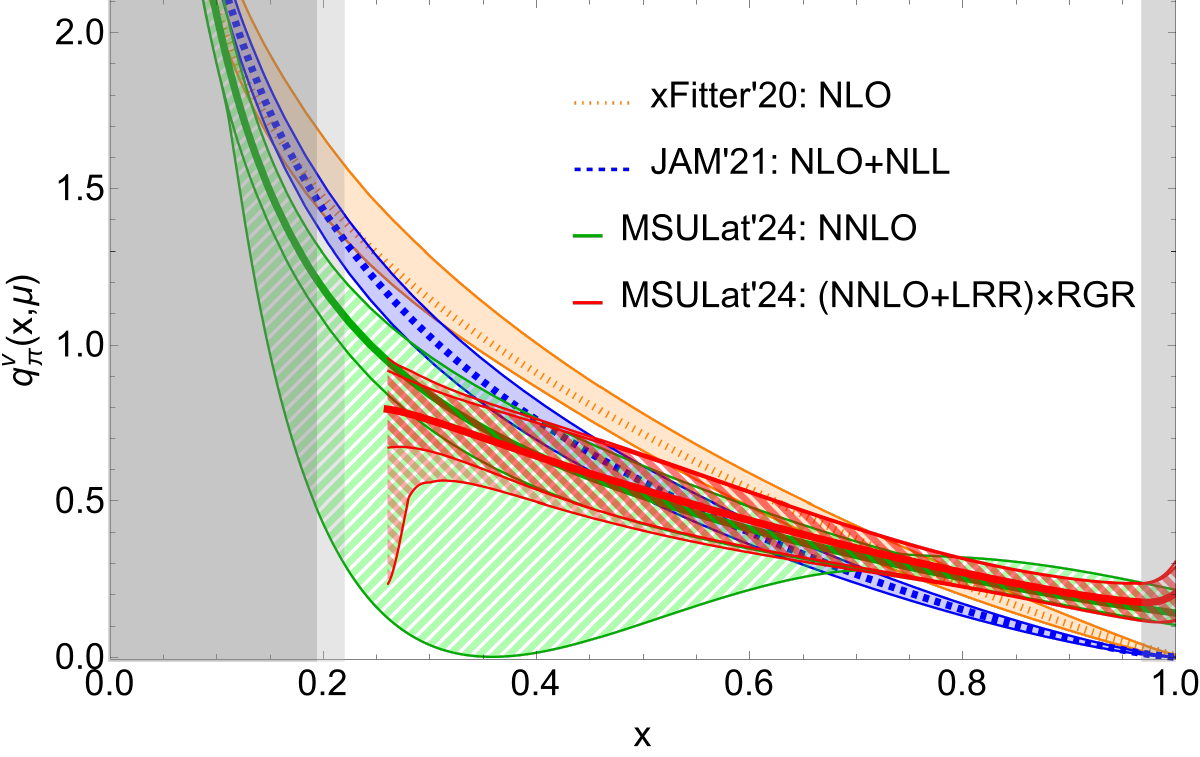}
    \caption{Comparison of our $x$-dependent PDFs (``\texttt{MSULat'24}") at \NN~(hatched green) and \NNLR~(hatched red) with global fits computed in Refs.~\cite{Novikov:2020snp} (``\texttt{xFitter'20}") at \N~(dotted orange) and \cite{Barry:2021osv} (``\texttt{JAM'21}") at \N~with next-to-leading logarithms (dashed blue). The \texttt{xFitter'20} results have been scaled so as to satisfy Eq.~(\ref{eq.Mellin0}).}
    \label{fig:PDF4}
\end{figure}

We compare our numerical results with those of global fits performed by the JAM~\cite{Barry:2021osv} and the xFitter~\cite{Novikov:2020snp} collaborations in Fig.~\ref{fig:PDF4}. We scale the results of the xFitter collaboration to match our convention of valence quark number conservation:
\begin{equation}\label{eq.Mellin0}
    \int^1_0\diff{}{x}q^{\rm v}_{\pi}(x,\mu)=1.
\end{equation}
This is also the same convention that the JAM collaboration uses.  
We focus on the JAM results with next-to-leading logs (NLL) since they offer a better systematic improvement of the valence quark PDF.
Our results include both statistical errors and combined statistical and systematic errors. We show our results at \NN~and \NNLR~to show the difference between two-loop matching with and without RGR and LRR improvements.
Our \NNLR~results agree within two sigma with the JAM results in the interval $x\approx[0.24,0.83]$ and with the xFitter results in the interval $x\approx[0.35, 0.95]$.
The difference we see in the mid-$x$ region is likely due to the different levels of improvement. 
Note that the JAM results~\cite{Barry:2021osv} computed with and without NLL also shows 1-2 sigmas difference within the same analysis frame; the latter one has better agreement with the xFitter results. 
Likely we are seeing difference due to the different systematic improvements from NLL, RGR and LRR.
In the large-$x$ region, our results are larger than those of JAM and xFitter who use the parametrization form of $x^a(1-x)^b$ to ensure the PDF goes to 0 at $x=1$, while ours does not. 
This same behavior has occurred in other LaMET framework calculations of the valence-quark pion PDF done by ANL/BNL group~\cite{Gao:2022iex} at NNLO level.
Overall, there is reasonable agreement between our numerical results, previous LaMET calculations and the global fits of JAM and xFitter.

\section{Conclusion}\label{sec.Conclusion}

In this paper we have computed the pion valence-quark PDF with physical quark mass at boost momentum $1.72$~GeV with the improvements of RGR and LRR. 
We use a physical pion mass ensemble generated by the MILC collaboration \cite{MILC:2010pul,MILC:2012znn,MILC:2015tqx} with lattice spacing $a\approx 0.09$~fm, and 2+1+1 flavors of highly improved staggered quarks in the sea and clover fermion for the valence sector. 
We compute and compare the LaMET matrix elements renormalized in the hybrid-ratio and hybrid-RI/MOM schemes as well as the corresponding $x$-dependent PDFs with matching performed at both one- and two-loops levels.

We report the impacts of different levels of improvement in the renormalization and matching from $\delta m$-only to implementation of RGR and LRR.
We found that the renormalized matrix elements in both the hybrid-RI/MOM and hybrid-ratio scheme are consistent with each other within the statistical errors, but the former have slightly higher central value across all the Wilson-line displacements we studied.
The systematic errors from scale variation in the renormalized matrix elements in the hybrid-RI/MOM and hybrid-ratio  schemes are greatly reduced by the simultaneous application of RGR and LRR 
 at one- and two-loop level respectively.  
However, the application of RGR on its own at either level increases the systematic errors, due to its enhancement of the renormalon ambiguity.
We found our pion valence-quark PDF in hybrid-RI/MOM and hybrid-ratio scheme to be consistent with each other within one sigma, both in the case of $\delta m$ only and with RGR and LRR improvement. 
Unfortunately, there are no results in the literature with which we can compare directly, since the only previously calculated NNLO pion valence-quark PDF was given by Ref.~\cite{Lin:2023gxz}, renormalized in the hybrid-ratio scheme, but these LaMET systematics were not included. 
In studying the $x$-dependent PDFs, we also show that there is an increase in systematic errors between \N\ and \NN\ but a decrease in systematic errors from \NLR\ to \NNLR\ and better convergence of the central values. 
This demonstrates that the inclusion of higher orders in both the renormalization and matching processes must be supplemented with an improved handling of the systematic errors if the results are to be made more precise. 
Our \NNLR\ result is the most reliable, since it includes the highest-order lightcone matching and accounts for both the large logarithms and the renormalon ambiguity. We also compare our results to global fits performed by the JAM and xFitter collaborations. Overall, we have reasonable agreement up to differences due to improvements from RGR and LRR.

\section*{Acknowledgments}
We thank the MILC Collaboration for sharing the lattices used to perform this study. We also thank Patrick Barry of the JAM collaboration and Ivan Novikov of the xFitter collaboration for sharing their respective $x$-dependent valence-quark PDFs at 2 GeV.
JH thanks Yushan Su and Rui Zhang for discussions and help with the LRR and RGR implementations.
The LQCD calculations were performed using the Chroma software suite~\cite{Edwards:2004sx}.
This research used resources of the National Energy Research Scientific Computing Center (NERSC), a DOE Office of Science User Facility supported by the Office of Science of the U.S. Department of Energy under Contract No.~DE-AC02-05CH11231 through ERCAP;
Advanced Cyberinfrastructure Coordination Ecosystem: Services \& Support (ACCESS) program~\cite{boerner2023access}, which is supported by National Science Foundation grants 2138259, 2138286, 2138307, 2137603, and 2138296;
the Extreme Science and Engineering Discovery Environment (XSEDE)~\cite{towns2014xsede}, which was supported by National Science Foundation grant number 1548562;
facilities of the USQCD Collaboration, which are funded by the Office of Science of the U.S. Department of Energy,
and supported in part by Michigan State University through computational resources provided by the Institute for Cyber-Enabled Research (iCER). 
The work of JH and HL are partially supported by the US National Science Foundation under grant PHY 1653405 ``CAREER: Constraining Parton Distribution Functions for New-Physics Searches'', grant PHY 2209424, and by the Research Corporation for Science Advancement through the Cottrell Scholar Award.

\bibliography{ref}
\end{document}